\documentclass[aip,
 amsmath,amssymb,
 reprint,%
]{revtex4-1}

\draft 

\usepackage{graphicx}
\usepackage{dcolumn}
\usepackage{bm}
\usepackage{color}

\usepackage[utf8]{inputenc}
\usepackage[T1]{fontenc}
\usepackage{mathptmx}

\begin{document}


\title{Screening current effect on the stress and strain distribution in REBCO high-field magnets: experimental verification and numerical analysis} 

\author{Yufan Yan}
 \affiliation{Department of Mechanical Engineering, Tsinghua University, Beijing 100084, China}
\author{Canjie Xin}
\affiliation{Institute of Modern Physics, Chinese Academy of Sciences, Lanzhou 730000, China}
\author{Yunfei Tan}
\affiliation{Wuhan National High Magnetic Field Center, Huazhong University of Science and Technology, Wuhan 430074, China}
\author{Timing Qu}
\affiliation{Department of Mechanical Engineering, Tsinghua University, Beijing 100084, China}
\email{tmqu@tsinghua.edu.cn}

\date{10 October 2019}

\begin{abstract}
Besides screening-current-induced magnetic fields (SCIF), the shielding effect in high-$T_c$ coated conductors also has an strong influence on its stress/strain distribution in a coil winding, especially during high-field operations. To demonstrate this phenomenon, a special experimental setup was designed. With an LTS background magnet and a small HTS insert coil, we were able to carry out direct observations on the hoop strains of a 10-mm wide REBCO sample. Measured data was compared against numerical solutions solved by electromagnetic models based on $T$-$A$ formulation and homogeneous mechanical models, showing good agreements. An analytical expression was proposed to estimate the maximum radial Lorentz force considering the shielding effect. Using the developed numerical models, we further studied the potential effects of two of the mostly investigated methods, which were formerly introduced to reduce SCIF, including multi-filamentary conductors and current sweep reversal (CSR) approach.
\end{abstract}

\maketitle 


REBCO (REBa$_2$Cu$_3$O$_x$) coated conductors, known as second-generation high temperature superconductors (2G HTS), are promising candidates for future high-field magnets\cite{Bonura2016}. Thanks to advancements in material design and manufacture, higher critical current, better in-field performance, and mechanical strength are being achieved\cite{Sundaram2016,Zhao2019}. Considering technical applicability and cost efficiency, the hybrid solution utilizing HTS magnets as insert field boosters is favored by a lot of researchers\cite{Barth2016}. Design and experiments on H800, an HTS insert magnet for 1.3 GHz NMR project at MIT, are underway\cite{Li2019}. Upon achieving center field of 27$\:$T in an LTS-HTS system, researchers at Institute of Electrical Engineering (IEE,CAS) launched a new project targeting a 30-T superconducting magnet for quantum oscillation application \cite{Liu2019}. At the National High Magnetic Field Laboratory (NHMFL), all superconducting user-magnet with a cold bore size of 34 mm reached the landmark full field of 32$\:$T\cite{Weijers2016}.

The influence of shielding currents on the magnetic fields of REBCO magnets is wildly recognized and studied\cite{Brandt1993,Yanagisawa2010}. It is confirmed by both experimental and numerical studies that the time drift and field distortion brought up by the screening-current-induced magnetic fields (SCIF) deteriorate the overall field quality\cite{Noguchi2019,Pardo2016}. Several approaches including filamentation of REBCO tapes and current sweep reversal have been proposed and tested to minimize the negative impacts\cite{Kajikawa2011,Yanagisawa2015}.

Recently, as is indicated by some studies, when taking the screening effect into account, the non-uniformly distributed currents in REBCO tapes may have a significant impact on the stress and strain distribution\cite{Li2019,Xia2019}, despite its high tolerance for tensile stress\cite{Iwasa2013}. In particular, in the pioneer experiment reaching 45.5-T with 14.4-T HTS insert, the one-side wrinkle pattern demonstrated in REBCO tapes also suggests the influence of screening currents\cite{Hahn2019}. But till now, validation experiments conducted in controlled environments are still missing.

\begin{figure}
\includegraphics[width=0.48\textwidth]{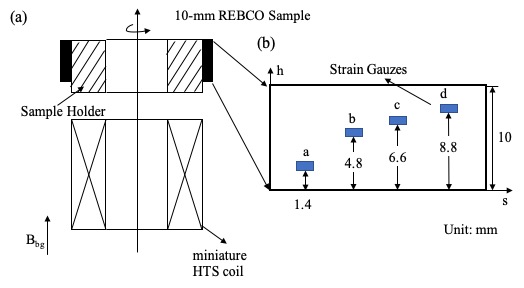}
\caption{\label{fig1}(a) Experimental setup for in-field strain test. A small HTS coil was co-axially assembled with the sample REBCO tape of 10-mm width, providing dynamic radial field. They were then inserted into an LTS magnet, generating stable parallel field for the sample. (b) Sensors were arranged and numbered a to d from left to right-hand side. Here $s$ and $h$ directions are not plotted in scale.}
\end{figure}

To demonstrate this phenomenon, we designed an experiment with simplified structures as illustrated in Fig.~\ref{fig1}. A single turn of REBCO tape with a width of 10$\:$mm, provided by Shanghai Superconductor, was prepared to serve as the test sample. We attached 4 low-temperature strain gauges (effective width of 1$\:$mm) in total, arranged in an oblique line as shown in Fig.~\ref{fig1}(b). Two ends of the sample were welded to the copper terminals on an arc-shaped sample holder with a radius of 17$\:$mm. The sample holder was made of G10, which has a much larger coefficient of thermal contraction comparing to REBCO tape, providing reasonable space for the contraction of the sample while maintaining the structure. A small HTS coil, with inner/outer diameter and winding height of 14$\:$mm, 34$\:$mm and 52$\:$mm, was coaxially aligned to generate the radial fields for the test sample. This whole assembly was then inserted into a low-temperature superconducting magnet with a bore size of 70$\:$mm, providing up to 9.0-T parallel field at the tape surface to magnify the strain response.

\begin{figure}
\includegraphics[width=0.48\textwidth]{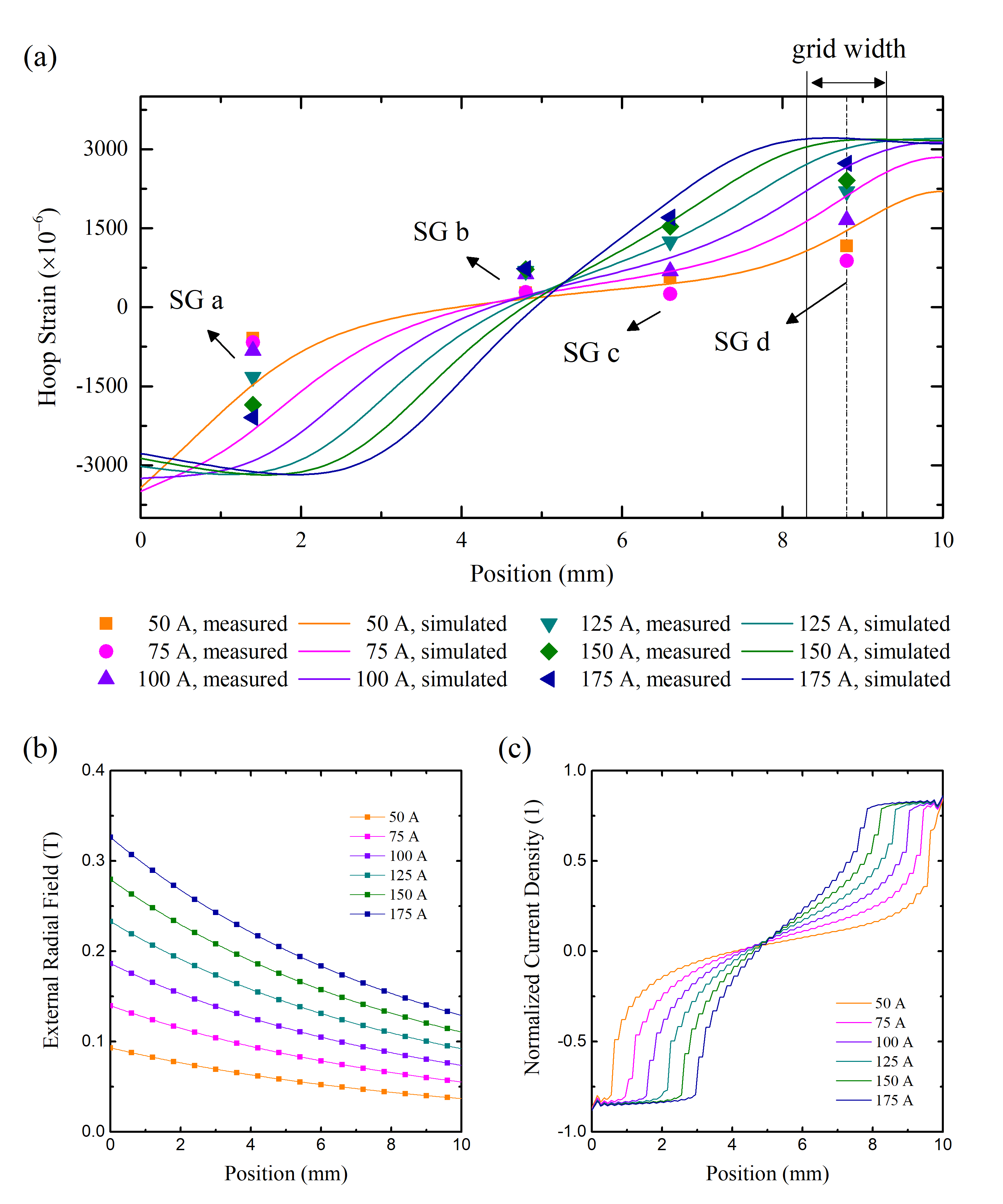}
\caption{\label{fig2}(a) Measured and calculated hoop strains from strain gauge SG a - SG d, plotted against their axial position in $h$ direction as defined in Fig.~\ref{fig1}. Grid width of SG d was marked for reference. (b) Radial external field profile at sample surface. (c) Calculated current distribution.}
\end{figure}

With this arrangement, the tested sample was staged in an environment similar to the outermost turn located on the axially outer side of a solenoid coil winding, whereas no transport current was applied. According to the following classic formula, which is generally used to estimate the hoop stress for high-field magnet design,
\begin{equation}
\label{eq1}
	\sigma = B_{\rm{z}} J_\varphi r\:,
\end{equation}
where $B_{\rm{z}}$ is the axial magnetic field for a solenoid coil, $J_\varphi$ is the averaged current density, and $r$ is the radius. Under the assumption of uniform current distribution, hoop stress in this test case should be zero. While, our experiments showed that as the coil was energized, tensile and compressive hoop strains were induced to the sample accordingly (see Fig.~\ref{fig2}(a)).

In order to understand its mechanism, numerical models were developed. For electromagnetic fields, we adopted the axisymmetric $T$-$A$ formulation\cite{Zhang2017,Liang2017}. The governing equations for the axially aligned superconducting sheet with thin-strip approximation is
\begin{equation}
\label{eq2}
	\frac{d}{dz}E_\varphi\left(\frac{dT}{dz}\right)=-\frac{\partial B_r}{\partial t}\:,
\end{equation}
where the state variable $T$ is the current vector potential, $J$ is the current density and $E$ is the electrical field. The magnetic field, $B$ then could be incorperated by using $A$ formulation embeded in commercial FEM software. The virtual thickness of the superconducting layer is set to be 1 $\mu$m. The superconducting property is modeled with $E$-$J$ power law with the n-index set as 16, and anisotropic Kim model fitted by experimental data provided by the manufacturer, where $J_{c0}$, $B_0$, $k$ and $\alpha$ are 5.3$\times 10^{11}~{\rm{A}}/{\rm{m}}^2$, 0.59 T, 9.1$\times 10^{-3}$ and 0.60, respectively.

Then radial Lorentz fasorce profiles are extracted and applied to the axisymmetric elastic model, expressed as
\begin{equation}
f_r = J_\varphi B_z\:,
\\
f_z = -J_\varphi B_r\:,
\end{equation}
where $B_r$ and $B_z$ is the radial and axial magnetic field. Here, the sample is simplied and modeled as four layers with their mechanical properties listed in Tab.~\ref{tab1}. In consistence with measured hoop strain, thermal contractions are excluded.

The measured and calculated hoop strains when HTS coil was energized with transport currents of 75$\:$A to 175$\:$A were plotted in Fig.~\ref{fig2}(a). Reasonable agreements could be obtained. On the axially outer side, tensile hoop strain up to +2800 microstrains was observed at sensor d with transport current of 175$\:$A. Contrarily, compressive strains were measured on the opposite side. While near the center, readings at strain gauge b showed minimal changes.

\begin{table}
\caption{\label{tab1}Material properties used in mechanical models.\cite{Ilin2015,Ekin2006}}
\begin{ruledtabular}
\begin{tabular}{ccccc}
&&Young's&Poisson's\\
Material&Thickness&Modulus&Ratio\\
\hline
REBCO & 1$\:\mu$m & 157$\:$GPa & 0.3\\
Copper & 22$\:\mu$m & 89$\:$GPa & 0.34\\
Hastelloy & 50$\:\mu$m & 228$\:$GPa & 0.307\\
\end{tabular}
\end{ruledtabular}
\end{table}

For comparison, radial external field at the sample surface and numerically simulated current distribution are shown in Fig.~\ref{fig2}(b) and (c). When subjected to the radial fields generated by the HTS coil, shielding currents were induced in the test sample, in opposite directions near the axially upper and lower side. The frontier of the critical region moved towards the center as the perpendicular fields increased. With the axial field predominately provided by the LTS background magnet, the radial Lorentz force, which contributed the most to the deformation of the sample, presented similar pattern to that of current distribution profile. Hence, along the axially outer rim of the sample, where radially outward Lorentz forces were expected, tensile deformation enlarged with increased external radial fields. While on the opposite side, it showed hoop strains in contraction accordingly.

\begin{figure}
\includegraphics[width=0.48\textwidth]{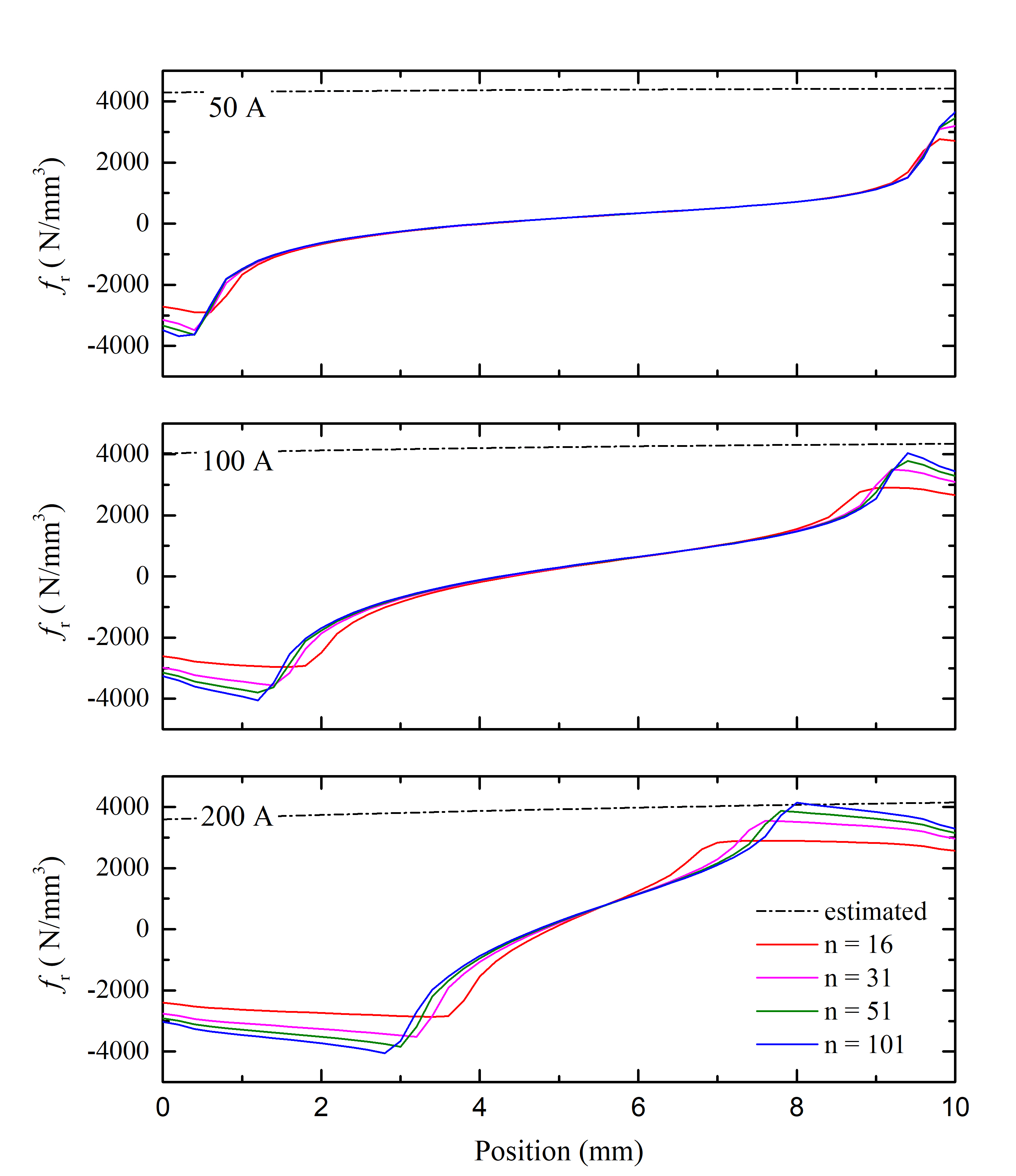}
\caption{\label{fig3} Radial Lorentz force calculated with n equals to 16, 31, 51 and 101 when the HTS coil was energized with transport current of 50$\:$A, 100$\:$A and 200$\:$A. Estimated values of maximum radial Lorentz force are plotted as well.}
\end{figure}

Without extra reinforcement techniques such as overbanding\cite{Qu2017}, Lorentz force in radially outward direction could cause most threat to the outer turns in a solenoid coil winding. Based on previous models and analyses, we propose the following analytical expression to estimate this component of Lorentz force considering screening current effect,
\begin{equation}
f_r \approx J_{c}(B_{\rm{r}},B_{\rm{z}})B_{\rm{z}}\:,
\label{eq4}
\end{equation}
where $J_{c}(B_{\rm{r}},B_{\rm{z}})$ is the field dependent critical current profile of the tape. The axial field $B_{\rm{z}}$ is composed of $B_{\rm{0}}$, which is contributed by background field magnet, and $B^*$, which was generated by the insert coil with uniform current distribution assumption and could be obtain by simple static models. In HTS insert magnet design, at the outer turns of a coil winding where the axial fields are dominated by the background magnet, $B_{\rm{z}}$ could be further replaced with $B_{\rm{0}}$.

This could be confirmed by theoretical models. Fig.~\ref{fig3} shows the radial Lorentz force profiles as n is set to be 16, 31, 51 and 101, with the upper bound predicted by the proposed Eq.~\ref{eq4}. By taking n-index in $E$-$J$ power law to an extreme, the solution conforms to that of the critical-state model, with the maximum current density approaching critical current density $J_{\rm{c}}$. These calculations also indicate that, practically, by lowering the value of n, the maximum radial Lorentz force could be reduced, as suggested in Ref.~14.

From Eq.~\ref{eq4}, the following conclusion could also be drawn. Although higher critical current density generally implies larger operation margin for current-carrying ability and thermal stability, it could lead to higher radial Lorentz force, limiting its tensile stress margin instead.

\begin{figure}
\includegraphics[width=0.48\textwidth]{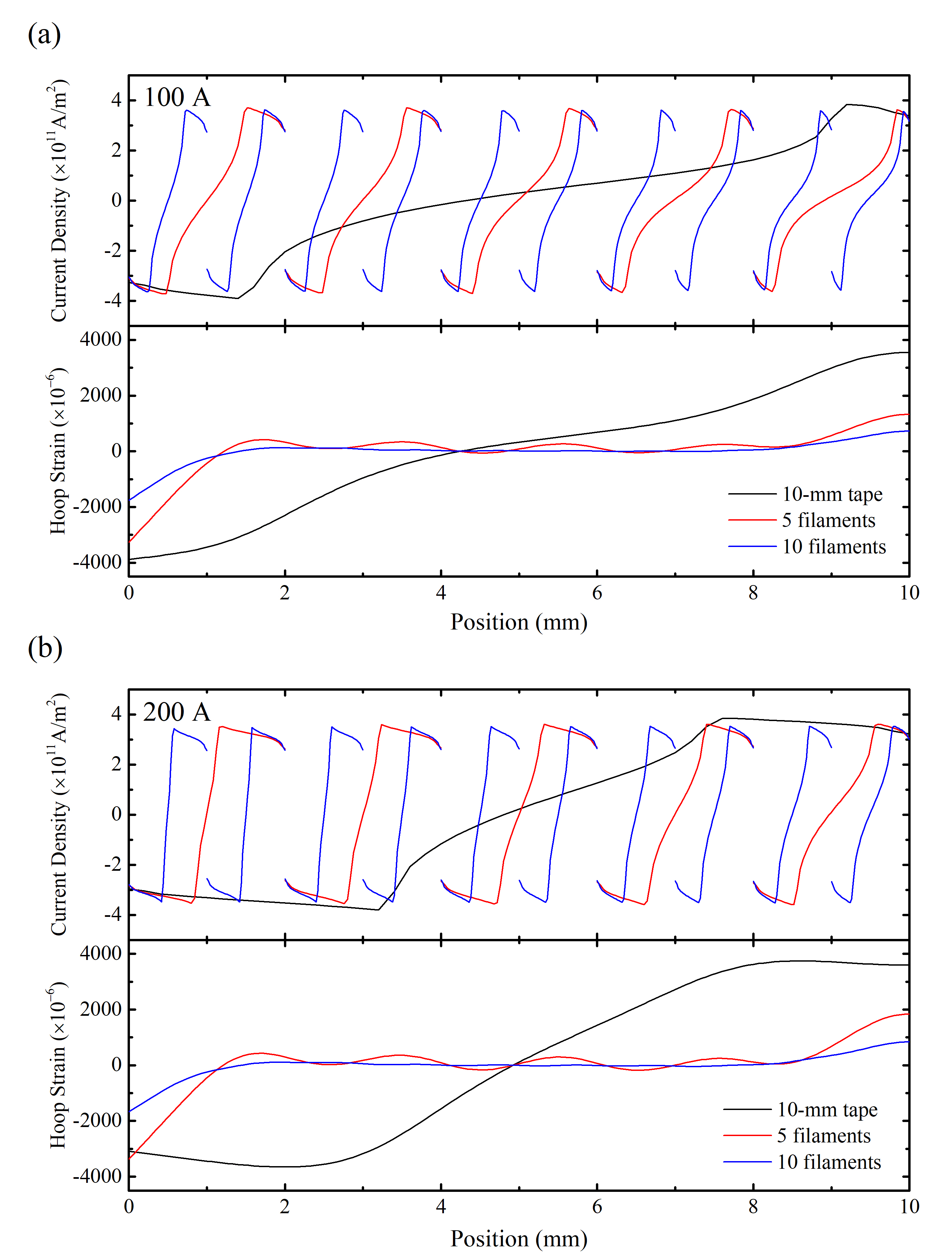}
\caption{\label{fig4} Calculated current density and hoop strain for samples with 1, 5 and 10 filaments, with transport current of 100$\:$A and 200$\:$A for HTS coil.}
\end{figure}

Multi-filamentary conductors and current sweep reversal (CSR) are two of the most wildly tested approaches to eliminate the SCIF in REBCO magnets\cite{Yanagisawa2015,Hwang2018}. Here we also investigated their effects on the stress and strain distribution using the numerical models we developed. For multi-filamentary REBCO conductors, we modelled samples with 5 and 10 filaments at transport current of 100$\:$A and 200$\:$A for HTS coil, while other parameters were consistent with previous models. As shown in Fig.~\ref{fig4}, by constraining shielding effect within filaments, electromagnetic forces were distributed equivalently. The tensile hoop strains were effectively mitigated. Specifically, at 200$\:$A, the maximum tensile hoop strain was reduced by factor of 0.51 and 0.77 for 5 and 10 filaments, respectively. But it should also be noted that this type of conductors are generally more prone to mechanical defects introduced during the striation process such as mechanical cutting and laser scribing\cite{Grilli2016}.

\begin{figure}
\includegraphics[width=0.48\textwidth]{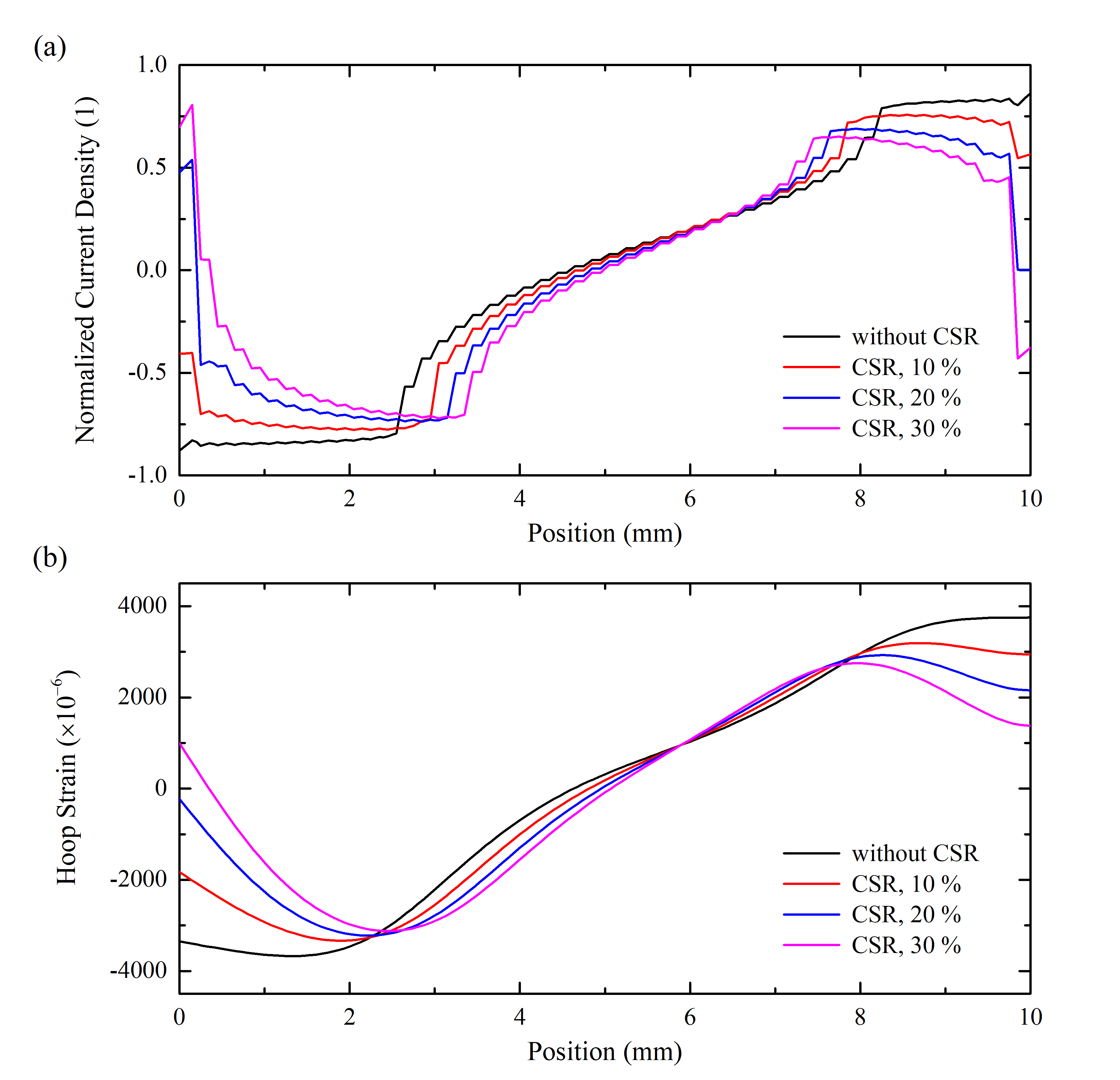}
\caption{\label{fig5}Calculated hoop strain after single reverse cycle with overshooting factors of 0, 10\%, 20\% and 30\% with target operating current of HTS coils at 150$\:$A.}
\end{figure}

For the method of current sweep reversal, we examined the cases of single reverse cycle with the overshooting factors of 10\%, 20\% and 30\%. Target operating current for HTS coil was set at 150$\:$A. Calculated normalized current distribution profiles and hoop strains were summaried in Fig.~\ref{fig5}. By overshooting and reverse process, screening currents along sample edges were reversed, and hoop strains at the side peviously in tension (right-hand side in Fig.~\ref{fig5}) could be lowered down. But with even higher overshooting factor, the opposite rim, which was in compression during normal loading process, could present tensile hoop strain in reverse.

In summary, we experimentally demonstrated the influence of screening current effect on the strain/stress distribution in REBCO coil winding. Numerical modelling with electromagnetic models and mechanical models confirmed this idea. Based on these analyses, we proposed an analytical expression to estimate the maximum radial Lorentz force, assisting future REBCO high-field magnet design. Finally, we carried out numerical studies on the effect of filamentized tapes and the method of current sweep reversal, which were previously developed for reducing screening current induced magnetic field. Both showed promising potentials in reducing the influence of screening current effect on the mechnical properties of REBCO magnets.\\

This work was supported by the Strategic Priority Research Program of Chinese Academy of Sciences under Grant No. XDB25000000 and National Natural Science Foundation of China (U1632276).

\bibliography{APL_submission}

\end{document}